\documentclass[%
 reprint,
 amsmath,amssymb,
 aps,
]{revtex4-2}
\usepackage{graphicx}
\usepackage{dcolumn}
\usepackage{bm}
\usepackage{subfigure}
\usepackage[dvipsnames]{xcolor}
\usepackage{amssymb,mathtools}
\usepackage{bigints}

\begin{document}

\preprint{APS/123-QED}

\title{Nonlocal characteristics and argand diagram of two-qubit gates }

\author{M. Karthick Selvan}
\email{karthick.selvan@yahoo.com}%

\author{S. Balakrishnan}%
 \email{physicsbalki@gmail.com}
\affiliation{Department of Physics, School of Advanced Sciences, Vellore Institute of Technology, Vellore - 632014, Tamilnadu, India.}%



\begin{abstract}
Nonlocal characteristics of a two-qubit gate are determined by its nonlocal part. The squared eigenvalues of the nonlocal part of a two-qubit gate exist on the unit circle in the complex plane. We show that two sets of chords, the chords connecting the squared eigenvalues with each other and those connecting a squared eigenvalue with the complex conjugate of others in the unit circle, can be used to describe the nonlocal characteristics of two-qubit gates. Lengths of both sets of chords are proportional to the amount of entanglement contained in certain pure states. The entangling power of a two-qubit gate can be expressed in terms of the squared lengths of the first set of chords. Similarly, we show that the gate typicality of a two-qubit gate can be expressed in terms of the squared lengths of the second set of chords and the linear entropy of a two-qubit gate can be expressed using the squared lengths of both sets of chords. Perfect entanglers are known to transform some product states into maximally entangled states. The convex hull of the squared eigenvalues of the nonlocal part of perfect entanglers contain the zero. We analyse the simplices containing the zero in the convex hull of the squared eigenvalues of the nonlocal part of perfect entanglers to construct a pair of orthonormal product states that can be transformed into maximally entangled states by the nonlocal part of perfect entanglers and divide the region of perfect entanglers in the Weyl chamber into three tetrahedral regions and eight bounding planes based on the uniqueness of the simplices containing the zero. 
\end{abstract}

\maketitle
\section{Introduction} 
Understanding the geometry and nonlocal characteristics of two-qubit gates is essential as their role in quantum computation is vital~\cite{Barenco1995,Barenco1995s,DiVincenzo1995}. Over two decades, two-qubit gates have been studied vastly. Nonlocal characteristics of two-qubit gates are invariant under local operations. A complete set of local invariants of two-qubit gates was obtained in~\cite{Makhlin2002}. The geometry of local equivalence classes of two-qubit gates, called as Weyl chamber, was studied in detail in~\cite{Zhang2003}. A measure for operator entanglement was introduced in~\cite{Zanardi2001}. Nonlocal characteristics of two-qubit gates were considered as resources for doing quantum information processing and their quantification was studied in~\cite{Nielsen2003}. Entangling power, a measure of the ability of two-qubit gates to generate entanglement, was defined in~\cite{Zanardi2000}; its expression in terms of Cartan co-ordinates~\cite{Rezakhani2004} and local invariants~\cite{Balakrishnan2010} were derived. Gate typicality as complementary to entangling power was introduced~\cite{Jonnadula2017} and its properties for two-qubit gates were studied~\cite{Jonnadula2020}. 

Despite many studies still, there remain unexplored ways to understand the nonlocal characteristics of two-qubit gates. Convex hull of the squared eigenvalues of the nonlocal part of two-qubit gates is either a $k$-simplex $(k \leq 2)$ or a quadrilateral inscribed on the unit circle in the complex plane. Recently, the chords, connecting the squared eigenvalues of the nonlocal part of two-qubit gates in their argand diagram, were shown to describe the ability of two-qubit gates to generate entangled states~\cite{Selvan2024}. This argand diagram was studied in regard to the condition for perfect entanglers~\cite{Zhang2003,Makhlin2002}, operational discrimination of the nonlocal part of a two-qubit gate from the nonlocal part of its adjoint~\cite{Chefles2005}, and simulation of perfect entanglers~\cite{Yu2010}. Perfect entanglers can transform some product states into maximally entangled states. Perfect entanglers occupy nearly $85\%$ of the invariant volume of two-qubit gates space~\cite{Watts2013,Musz2013}. Perfect entanglers are used as entangling basis gate for doing quantum computation~\cite{Chow2011,Debnath2016,Huang2023}.

Entangling power of a two-qubit gate was shown to be proportional to the mean squared length of the six chords connecting the squared eigenvalues of the nonlocal part of the two-qubit gate in its argand diagram~\cite{Selvan2024}. For each chord connecting a pair of squared eigenvalues, there is a chord connecting one of the squared eigenvalue with the complex conjugate of the other. In this paper, we show that the gate typicality of a two-qubit gate is proportional to the mean squared length of the chords connecting a squared eigenvalue of the nonlocal part with the complex conjugate of other three. We also show that linear entropy quantifying operator entanglement of two-qubit gates~\cite{Zanardi2001} can be described using the squared lengths of both sets of chords. 

For a two-qubit gate to transform a product state into maximally entangled state, the convex hull of the squared eigenvalues of its nonlocal part should contain the zero~\cite{Zhang2003,Makhlin2002}. This condition only allows to distinguish the region of perfect entanglers from non-perfect entanglers in the Weyl chamber [FIG.~\ref{Weyl}]. In this paper, we discuss the construction of a pair of orthonormal product states that can be transformed into maximally entangled states by the nonlocal part of a perfect entangler using the simplices in the convex hull of the squared eigenvalues of the nonlocal part. This allows to divide the region of perfect entanglers in the Weyl chamber further into three tetrahedral regions and eight bounding planes based on the simplices containing the zero.  

The convex hull of the squared eigenvalues of the nonlocal part of perfect entanglers existing on the three faces of the Weyl chamber is a triangle without any edge passing through the zero. The perfect entanglers with at least one chord passing through the zero are represented by five planes; three of them separate perfect entanglers from non-perfect entanglers and the remaining two divide the region of perfect entanglers without any chord passing through the zero into three tetrahedral regions. We derive three sufficient conditions, one for each tetrahedron of perfect entanglers without any chord passing through the zero, for the existence of a pair of orthonormal product states that can be transformed into maximally entangled states. These conditions are satisfied by the perfect entanglers on $c_3 = 0$ and $c_1 = \pi/2$ planes of the Weyl chamber. 

This paper is organized as follows. In section II, we give an overview of the nonlocal part of two-qubit gates and their aragnd diagram. In section III, we discuss the quantification of nonlocal characteristics of two-qubit gates using the chords in the argand diagram of the squared eigenvalues of the nonlocal part of two-qubit gates. In section IV, we discuss the construction of a pair of orthonormal product states that can be transformed into maximally entangled states by the nonlocal part of perfect entanglers using the simplices in the argand diagram of perfect entanglers and show that the region of perfect entanglers in the Weyl chamber can be divided further based on the uniqueness of the simplices containing the zero. In section V, we provide the conclusion. 

\section{Nonlocal part of two-qubit gates and argand diagram}
\begin{figure}
\includegraphics[scale=0.675]{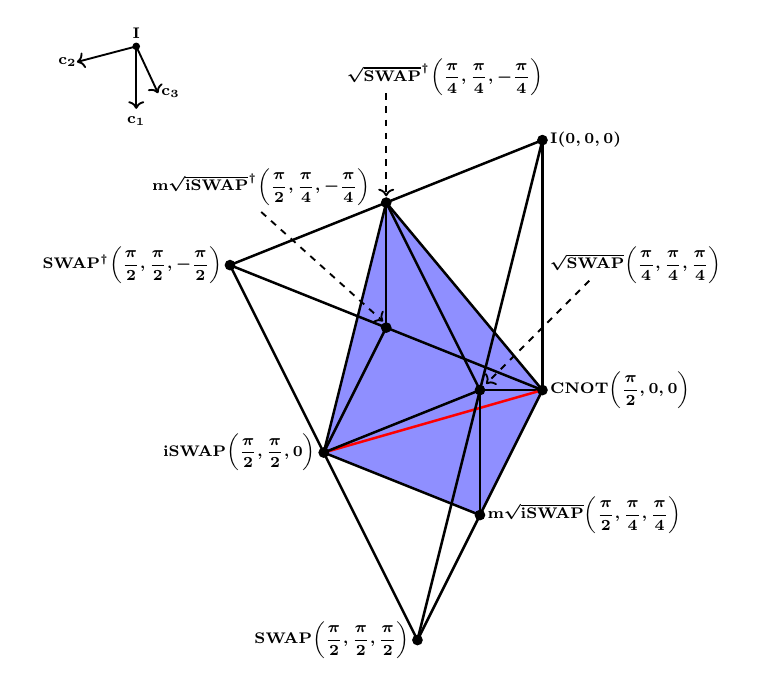} 
\caption{{Tetrahedron with vertices $\textnormal{I}(0,0,0)$, $\textnormal{CNOT}(\pi/2,0,0)$, $\textnormal{SWAP}(\pi/2, \pi/2, \pi/2)$, and $\textnormal{SWAP}^{\dagger}(\pi/2, \pi/2, -\pi/2)$. Orientation of Cartan co-ordinates are shown in the upper left corner. All the points of tetrahedron satisfy the condition: $\pi/2 \geq c_1 \geq c_2 \geq \vert c_3 \vert \geq 0$. Marked points are labelled by a gate representing the local equivalence class along with its Cartan co-ordinates. In the labels, $\textnormal{m}\sqrt{\textnormal{iSWAP}}$, is the mirror gate~\cite{Selvan2023} of $\sqrt{\textnormal{iSWAP}}$. Each local equivalence class is uniquely represented by a point of tetrahedron. However, as an exception, the points $(\pi/2, c_2, c_3)$ and $(\pi/2, c_2, -c_3)$ for given $c_2$ and $c_3$ represent the same local equivalence class. For example, the points $\textnormal{SWAP}(\pi/2, \pi/2, \pi/2)$, and $\textnormal{SWAP}^{\dagger}(\pi/2, \pi/2, -\pi/2)$ represent the same local equivalence class. Region coloured in blue is the region of perfect entanglers. The line in red colour represent special perfect entanglers (SPEs).}}
\label{Weyl}
\end{figure}

A two-qubit gate $U \in \textnormal{SU}(4)$ can be decomposed as follows~\cite{Zhang2003}. 

\begin{equation}\label{kUk}
U = L_1 U_d(c_1,c_2,c_3)L_2,
\end{equation}

where $L_1,~L_2 \in \textnormal{SU}(2) \otimes \textnormal{SU}(2)$ are local parts of $U$ and $U_d(c_1, c_2, c_3)$ is the nonlocal part of $U$. Two-qubit gates having the same nonlocal part but different local parts form a local equivalence class and each local equivalence class of two-qubit gates is geometrically represented as a point of tetrahedron shown in FIG.~\ref{Weyl}. This tetrahedron is also called as Weyl chamber~\cite{Zhang2003}. 

Nonlocal characteristics of $U$ is determined by its nonlocal part which can be written as  
\begin{equation}\label{NL}
U_d(c_1, c_2, c_3) = e^{iH/2}
\end{equation}
with 
\begin{equation}\label{Ham}
H = \sum_{j=1}^3 c_j \left(\sigma_j \otimes \sigma_j \right),
\end{equation}
where $c_1$, $c_2$, {and} $c_3$ are called Cartan co-ordinates. Eigenvalues of $H$ are given by 

\begin{eqnarray}
h_1 = c_1 -c_2 + c_3, \nonumber
\end{eqnarray}
\begin{eqnarray}
h_2 = c_1 +c_2 - c_3, \nonumber
\end{eqnarray}
\begin{eqnarray}
h_3 = -c_1 -c_2 - c_3, \nonumber
\end{eqnarray}

and

\begin{eqnarray}\label{EVHam}
h_4 = -c_1 +c_2 + c_3.
\end{eqnarray}

Eigendecomposition of the nonlocal part of $U$ can be written as shown below.   

\begin{equation}\label{NLM}
U_d(c_1,c_2,c_3) = e^{ih_3/2} \sum_{j=1}^4 \left[ e^{i(h_j-h_3)/2} \vert \Psi_j \rangle \langle \Psi_j \vert \right],
\end{equation}

where the column vectors $\vert \Psi_j \rangle$ form the eigen basis of $U_d$. It is also known as magic basis~\cite{Zhang2003} and they are expressed as $\vert \Psi_1 \rangle = \frac{1}{\sqrt{2}} \left[ \vert 00 \rangle + \vert 11 \rangle \right]$, $\vert \Psi_2 \rangle = \frac{i}{\sqrt{2}} \left[ \vert 01 \rangle + \vert 10 \rangle \right]$, $\vert \Psi_3 \rangle = \frac{1}{\sqrt{2}} \left[ \vert 01 \rangle - \vert 10 \rangle \right]$, and $\vert \Psi_4 \rangle = \frac{i}{\sqrt{2}} \left[ \vert 00 \rangle - \vert 11 \rangle \right]$. 

\begin{figure}
\centering
\includegraphics[scale=0.5]{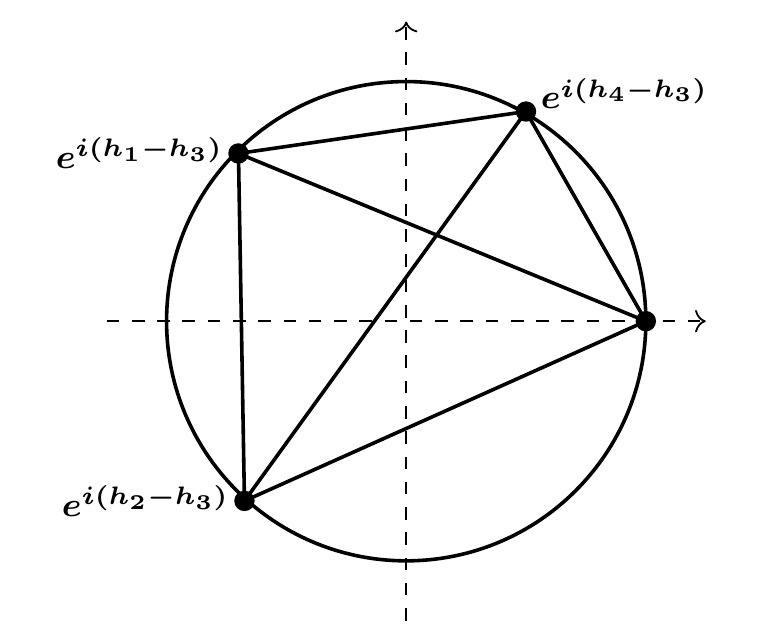} 
\caption{Argand diagram of squared eigenvalues of nonlocal part of a perfect entangler.}
\label{AD1}
\end{figure}

The squared eigenvalues of $U_d$ are on a circle of unit radius in the complex plane and they are pairwise connected by six chords forming a quadrilateral as shown in FIG.~\ref{AD1}. We refer to this argand diagram of squared eigenvalues of $U_d$ as the argand diagram of $U$. 

For perfect entanglers, there exist product states that can be transformed into maximally entangled states and the convex hull of squared eigenvalues of $U_d$ should contain zero as shown in FIG.~\ref{AD1}. The Cartan co-ordinates of perfect entanglers satisfy the following conditions~\cite{Selvan2024a}. 

\begin{eqnarray}
c_1 + c_2 \geq \frac{\pi}{2}, \nonumber
\end{eqnarray}
\begin{eqnarray}\label{PE}
c_2 \pm c_3 \leq \frac{\pi}{2}.
\end{eqnarray}

The region of perfect entanglers in the Weyl chamber (coloured in blue) is shown in FIG.~\ref{Weyl}. Special perfect entanglers which are perfect entanglers with maximum entangling power~\cite{Rezakhani2004} are represented by the red colour line in FIG.~\ref{Weyl}.

\section{Quantification of nonlocal characteristics using chords}
Like two-qubit states, two-qubit gates also possess entanglement and can be quantified using linear entropy~\cite{Zanardi2001,Wang2002}. Two-qubit gates represented by $c_3$ axis of the Weyl chamber have {a} maximum value of linear entropy~\cite{Balakrishnan2011}. Entangling power is another nonlocal measure of two-qubit gates that quantifies the average entanglement generated over {the} uniform distribution of product states~\cite{Zanardi2000,Rezakhani2004}. Gate typicality is a nonlocal measure that is complementary to entangling power~\cite{Jonnadula2017,Jonnadula2020}. Both entangling power $(e_p)$ and gate typicality $(g_t)$ together describe the linear entropy $(L)$ of two-qubit gates. The relationship between them can be written as follows.  

\begin{equation}\label{LE}
L=\frac{3}{8} \left[3e_p+g_t \right].
\end{equation}

Recently, the entangling power of a two-qubit gate was shown to be proportional to the mean squared length of the chords connecting the squared eigenvalues of its nonlocal part in the argand diagram~\cite{Selvan2024}. Entangling power can be written in terms of squared chord lengths as follows. 

\begin{equation}\label{EPC}
e_p = \frac{1}{72}\sum_{j=1}^3\sum_{k > j}\vert e^{ih_j} - e^{ih_k} \vert^2.
\end{equation}

\begin{figure}
\includegraphics[scale=0.5]{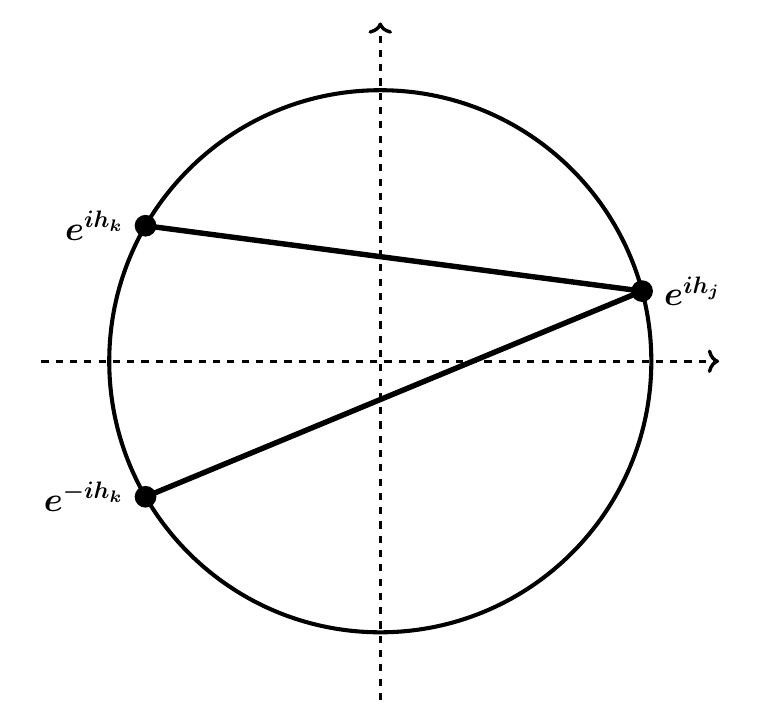}
\caption{Diagram showing two chords from $e^{ih_j}$ one connected to $e^{ih_k}$ and another to $e^{-ih_k}$.}
\label{EPGTC}
\end{figure}

Entangling power of a two-qubit gate $U$ can also be expressed as {a} symmetric combination of $L(U)$ and $L(US)-L(S)$ as follows~\cite{Wang2002,Balakrishnan2011,Jonnadula2017}. 

\begin{equation}\label{EPL}
e_p(U) = \frac{4}{9} \left( L(U) + \left[L(US) -L(S) \right] \right),
\end{equation}

where $L(U)$, $L(US)$, and $L(S)$ are linear entropies of the two-qubit gate $U$, its mirror gate, and SWAP gate respectively. 

Gate typicality of a two-qubit gate $U$ was defined as {an} antisymmetric combination of $L(U)$ and $L(US)-L(S)$~\cite{Jonnadula2017}.

\begin{equation}\label{GTL}
g_t(U) = \frac{4}{3} \left( L(U) - \left[L(US) - L(S) \right] \right). 
\end{equation}

Its expression in terms of the Cartan co-ordinates is given by 

\begin{equation}\label{GTCC}
g_t = 1 - \dfrac{1}{3} \left[\cos(2c_1) + \cos(2c_2) + \cos(2c_3) \right].
\end{equation}

It can be noted that six times the right-hand side of Eq.~\ref{EPC} with $e^{ih_k}$ replaced by its complex conjugate provides the expression of gate typicality of two-qubit gates. That is, 

\begin{equation}\label{GTC}
g_t = \frac{1}{12}\sum_{j=1}^3\sum_{k > j}\vert e^{ih_j} - e^{-ih_k} \vert^2. 
\end{equation}

The length of the chord connecting $e^{ih_j}$ and $e^{ih_k}$ [FIG.~\ref{EPGTC}] was shown to be proportional to the amount of entanglement contained in the states $\left[U_d \vert \Psi_j \rangle \pm i U_d \vert \Psi_k \rangle \right]/\sqrt{2}$~\cite{Selvan2024}. It can be verified that the length of the chord connecting $e^{ih_j}$ and $e^{-ih_k}$ is proportional to the amount of entanglement of the states $\left[U_d \vert \Psi_j \rangle \pm i U_d^{\dagger} \vert \Psi_k \rangle \right]/\sqrt{2}$. This replacement of $e^{ih_k}$ by its complex conjugate in Eq.~\ref{EPC} is similar to replacing the symmetric combination of $L(U)$ and $L(US)-L(S)$ in the definition of entangling power by its antisymmetric combination to define gate typicality. It has to be noted that if any chord describing the entangling power is parallel to the imaginary axis, then the length of the corresponding chord describing the gate typicality is zero. Similarly, if the length of a chord describing the entangling power is zero, then the corresponding chord describing the gate typicality is parallel to the imaginary axis. 

It can be verified that gate typicality can also be described using only three chords connecting any one of the squared eigenvalues with the complex conjugate of the remaining three squared eigenvalues as shown below.  

\begin{equation}\label{GTCA}
g_t = \frac{1}{6} \sum_{k \neq j} \vert e^{ih_j} - e^{-ih_k} \vert^2,~~\textnormal{for}~\textnormal{any}~j.
\end{equation}

However, the expression of gate typicality given in Eq.~\ref{GTC} has the same form as the expression of entangling power. Hence, it is useful to describe the difference between the mathematical definitions of entangling power and gate typicality.  

Since both the entangling power and gate typicality of a two-qubit gate can be calculated using the chords present in their argand diagram, the linear entropy (Eq.~\ref{LE}) describing the operator entanglement of the two-qubit gate can also be calculated using the chords as follows. 

\begin{equation}\label{LEC}
L = \frac{1}{64} \sum_{j=1}^3 \sum_{k > j} \left[\vert e^{ih_j} - e^{ih_k} \vert^2 + 2 \vert e^{ih_j} - e^{-ih_k} \vert^2 \right]
\end{equation}

\section{A pair of orthonormal product states to maximally entangled states}

In this section, we discuss the construction of a pair of orthonormal product states that can be transformed into maximally entangled states by the nonlocal part of a perfect entangler using the simplices formed by the squared eigenvalues of the nonlocal part of the perfect entangler. For a perfect entangler, if the chord connecting $e^{ih_j}$ and $e^{ih_k}$ (for $j,k \in \{1,2,3,4\}$ and $j \neq k$) passes through the zero, then the orthonormal pair of product states $\vert \Phi_{\pm} \rangle = \left[ \vert \Psi_j \rangle \pm i \vert \Psi_k \rangle \right]/\sqrt{2}$ can be converted into maximally entangled states by the nonlocal part of the perfect entangler~\cite{Selvan2024}. First, we identify the region of perfect entanglers, with at least one chord passing through the zero, in the Weyl chamber.

\subsection{Perfect entanglers with at least a chord passing through the zero}

\begin{figure}
\begin{tabular}{c}
\includegraphics[scale=0.65]{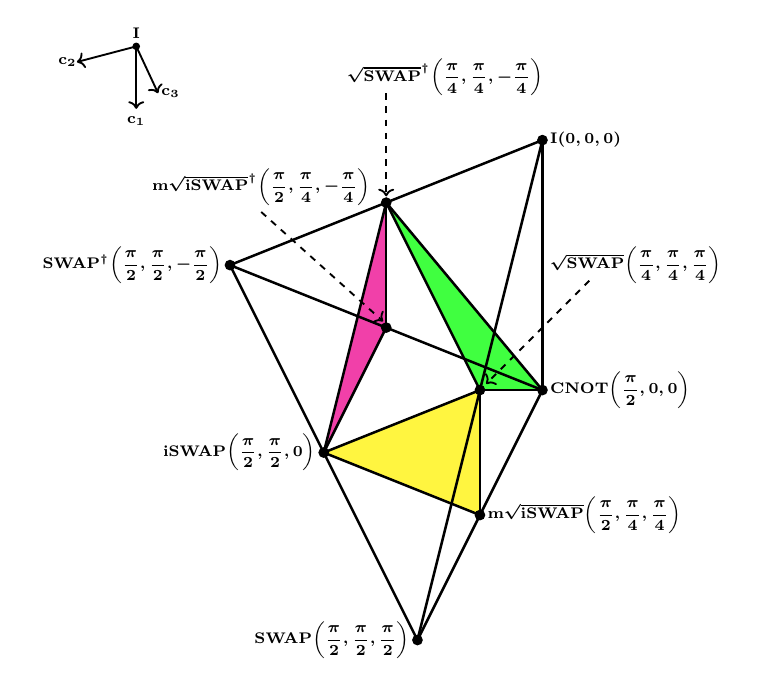} \\ (a) \\ \includegraphics[scale=0.65]{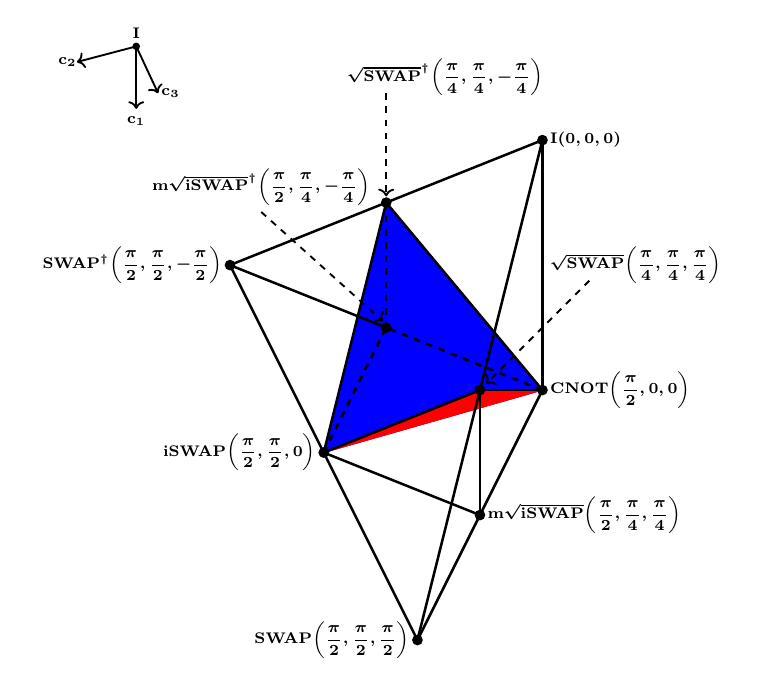} \\ (b)
\end{tabular}
\caption{Regions of perfect entanglers with at least one chord passing through the zero. In the subfigure (a) the planes coloured in magenta, green, and yellow are $\vert h_1 - h_2 \vert = \pi$, $\vert h_2 - h_3 \vert = \pi$, and $\vert h_3 - h_4 \vert = \pi$ planes respectively. In the subfigure (b), the planes coloured in red and blue are $\vert h_1 - h_3 \vert = \pi$ and $\vert h_2 - h_4 \vert = \pi$ planes respectively.}
\label{180C}
\end{figure}

The chord connecting $e^{ih_j}$ and $e^{ih_k}$ passes through {the} zero only if $\vert h_j - h_k \vert = \pi$. It can be verified that the equations $\vert h_j - h_{j+1} \vert = \pi$ with $j=1,2,3$ describe the three boundary planes separating perfect entanglers from {non-perfect} entanglers inside the Weyl chamber [Eq.~\ref{PE}]. These three planes are shown in FIG.~\ref{180C}a. Similarly, the equations $\vert h_j - h_{j+2} \vert = \pi$ with $j=1,2$ describe the planes $c_1 \pm c_3 = \pi/2$ [FIG.~\ref{180C}b]. {They belong to the reflecting planes describing the mirror operation~\cite{Selvan2023}}. The equation $\vert h_1 - h_4 \vert = \pi$ which can be rewritten as $c_1 - c_2 = \pi/2$ is satisfied only by the point representing CNOT equivalence class {with Cartan co-ordinates $(\pi/2,0,0)$}. Thus only the gates represented by the five planes shown in FIG.~\ref{180C}a and \ref{180C}b have at least one chord passing through {the} zero and hence they can transform a pair of orthonormal product states into a pair of orthonormal maximally entangled states. 

\begin{figure}[h]
\includegraphics[scale=0.675]{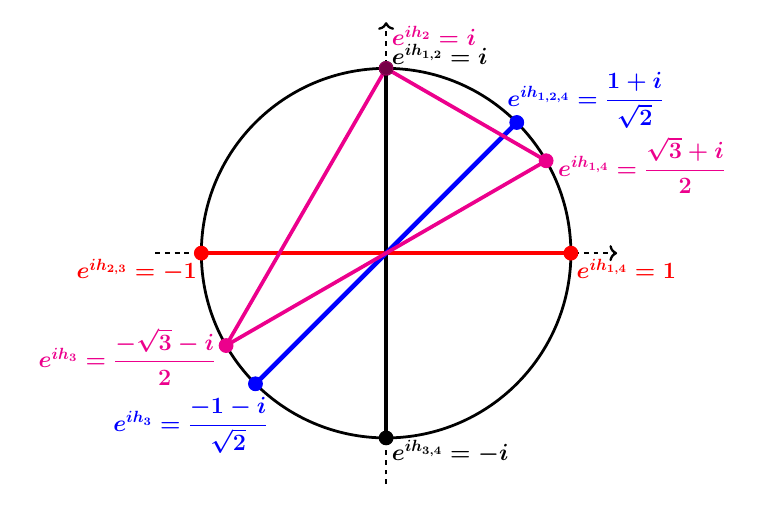}
\caption{Argand diagrams of the nonlocal parts of CNOT (black), iSWAP (red), $\sqrt{\textnormal{SWAP}}$ (blue) and an equivalence class with Cartan co-ordinates $(\pi/3, \pi/3, \pi/6)$ (magenta).}
\label{Example}
\end{figure}

All the perfect entanglers represented by a specific plane have a unique argand diagram with a chord, connecting a specific pair of squared eigenvalues, passing through the zero. However, there are gates present in more than one plane. SPEs are at the intersection of $\vert h_j - h_{j+2} \vert = \pi~(j=1,2)$ planes [FIG.~\ref{180C}b]. Hence, for SPEs, two chords connecting diametrically opposite points $(e^{ih_j}~\textnormal{and}~e^{ih_{j+2}}~\textnormal{with}~j=1,2)$ pass through the zero and they can transform orthonormal product basis into orthonormal maximally entangled basis~\cite{Selvan2024,Rezakhani2004}. Among SPEs, the point representing iSWAP equivalence class is also present in $\vert h_j - h_{j+1} \vert = \pi~\textnormal{with}~j=1,3$ planes [FIG.~\ref{180C}a]. Similarly, the point representing CNOT equivalence class also satisfies the conditions, $\vert h_2 - h_3 \vert = \pi$ and $\vert h_1 - h_4 \vert = \pi$. Hence, these two equivalence classes have argand diagrams with four chords passing through {the} zero~\cite{Selvan2024}. The argand diagram of nonlocal part of CNOT (black in colour) and iSWAP (red in colour) equivalence classes are shown in FIG.~\ref{Example}.

There are also local equivalence classes at the intersection of a plane shown in FIG.~\ref{180C}a and another plane shown in FIG.~\ref{180C}b. We have four such intersections between the following pairs of planes: $\{\vert h_3 - h_4 \vert = \pi, \vert h_1 - h_3 \vert = \pi\}$, $\{\vert h_2 - h_3 \vert = \pi, \vert h_1 - h_3 \vert = \pi\}$, $\{\vert h_1 - h_2 \vert = \pi, \vert h_2 - h_4 \vert = \pi\}$, and $\{\vert h_2 - h_3 \vert = \pi, \vert h_2 - h_4 \vert = \pi\}$. For these equivalence classes, two squared eigenvalues coincide. For example, for the local equivalence classes at the intersection of $\vert h_3 - h_4 \vert = \pi$ (the yellow coloured plane in FIG.~\ref{180C}a) and $\vert h_1 - h_3 \vert = \pi$ (the red coloured plane in FIG.~\ref{180C}b) planes, $e^{ih_1}$ coincides with $e^{ih_4}$. Hence, for all the local equivalence classes at the intersection of the pairs of planes mentioned above, the convex hull of squared eigenvalues is a triangle with one edge passing through the zero. The argand diagram of the nonlocal part of the local equivalence class with Cartan co-ordinates $(\pi/3, \pi/3, \pi/6)$, which is at the intersection of $\vert h_3 - h_4 \vert = \pi$ and $\vert h_1 - h_3 \vert = \pi$ planes, is shown in FIG.~\ref{Example} (magenta in colour).    

The point representing $\sqrt{\textnormal{SWAP}}$ equivalence class is common to $\vert h_3 - h_4 \vert = \pi$, $\vert h_1 - h_3 \vert = \pi$, and $\vert h_2 - h_3 \vert = \pi$ planes. Hence, in the argand diagram of $\sqrt{\textnormal{SWAP}}$ equivalence class, three chords from the point $e^{ih_3}$ pass through the zero. It implies that the remaining three points $e^{ih_1}$, $e^{ih_2}$, and $e^{ih_4}$ coincide with each other. 
The argand diagram is shown in FIG.~\ref{Example} (blue in colour). Similarly, it can be verified that the argand diagram of $\sqrt{\textnormal{SWAP}}^{\dagger}$ equivalence class has only three chords from the point $e^{ih_2}$ passing through {the} zero. Thus for the gates belonging to these two local equivalence classes there exist three different pairs of orthonormal product states that can be transformed into pairs of orthonormal maximally entangled states. 

\begin{figure}[h]
\includegraphics[scale=0.45]{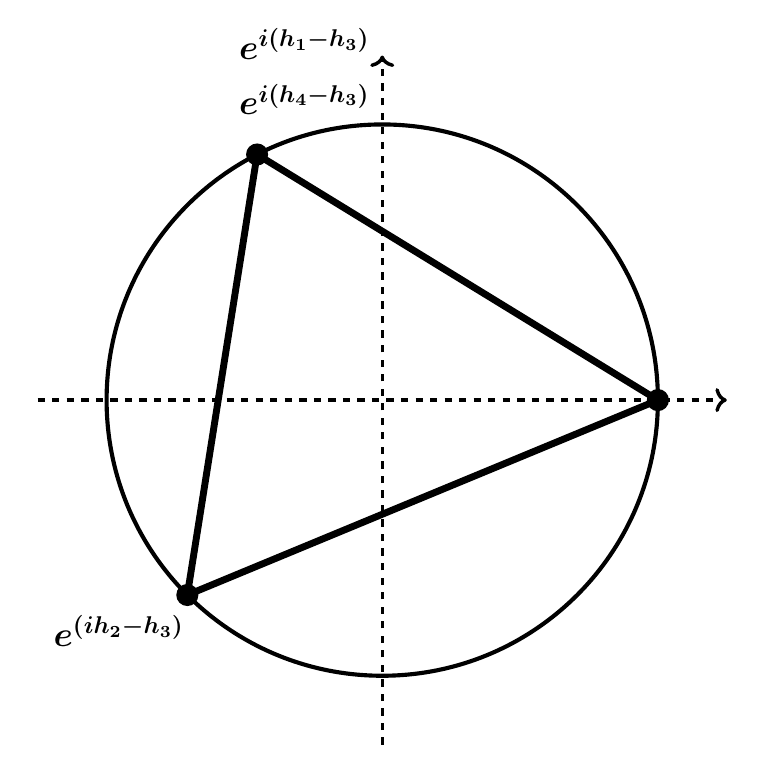} 
\caption{Typical argand diagram of the nonlocal part of a perfect entangler on $c_1=c_2$ plane of the Weyl chamber.}
\label{Triangle}
\end{figure}

\subsection{Perfect entanglers without any chord passing through the zero}
Now, we consider perfect entanglers without any chord passing through the zero in their argand diagram. First, we consider perfect entanglers existing on the faces of the Weyl chamber. For all the perfect entanglers existing on the faces of the Weyl chamber, the convex hull of the squared eigenvalues of their nonlocal part is a triangle. On $c_1 = c_2$ plane, perfect entanglers without any chord passing through the zero are contained within the triangular region bounded by $c_1 \pm c_3 = \pi/2$ and $c_1 = \pi/4$ lines [FIG.~\ref{180C}a]. For all these perfect entanglers, $e^{ih_1}$ coincides with $e^{ih_4}$  and the convex hull of the squared eigenvalues of their nonlocal part is a triangle as shown in FIG.~\ref{Triangle}. For the perfect entangler associated with the argand diagram, shown in FIG.~\ref{Triangle}, zero can be uniquely expressed as 

\begin{equation}\label{C1C2}
\vert \phi_3 \vert^2 e^{ih_3} + \vert \phi_1 \vert^2 e^{ih_1} + \vert \phi_2 \vert^2 e^{ih_2} = 0
\end{equation}

for some co-ordinates $\vert \phi_3 \vert^2$, $\vert \phi_1 \vert^2$, and $\vert \phi_2 \vert^2$ such that $\vert \phi_3 \vert^2 + \vert \phi_1 \vert^2 + \vert \phi_2 \vert^2 = 1$. From Eq.~\ref{C1C2}, it is possible to construct the following two product states~\cite{Zhang2003}

\begin{equation*}
\vert \Phi_1 \rangle = \vert \phi_3 \vert e^{-i(a\pi + h_3/2)} \vert \Psi_3 \rangle + \vert \phi_2 \vert e^{-i(b\pi + h_2/2)} \vert \Psi_2 \rangle 
\end{equation*} 
\begin{equation}{\label{TProd1}}
+ \vert \phi_1 \vert e^{-i(c\pi + h_1/2)} \vert \Psi_1 \rangle,
\end{equation}
and 
\begin{equation*}
\vert \Phi_2 \rangle = \vert \phi_3 \vert e^{-i(d\pi + h_3/2)} \vert \Psi_3 \rangle + \vert \phi_2 \vert e^{-i(f\pi + h_2/2)} \vert \Psi_2 \rangle 
\end{equation*} 
\begin{equation}{\label{TProd2}}
+ \vert \phi_1 \vert e^{-i(g\pi + h_1/2)} \left[ p \vert \Psi_1 \rangle + \sqrt{1-p^2} \vert \Psi_4 \rangle \right]
\end{equation}

for some integers $a$, $b$, $c$, $d$, $f$, and $g$ and $p \in [-1,1]$. These two product states can be converted into maximally entangled states by the nonlocal part of the perfect entangler associated with the argand diagram shown in FIG.~\ref{Triangle}. If the two product states, $\vert \Phi_1 \rangle$ and $\vert \Phi_2 \rangle$, are orthogonal to each other then the following condition is satisfied 

\begin{equation}{\label{TOrtho1}}
\vert \phi_3 \vert^2 e^{i(d-a)\pi} + \vert \phi_2 \vert^2 e^{i(f-b)\pi} + p \vert \phi_1 \vert^2 e^{i(g-c)\pi} = 0
\end{equation}

The weight $\vert \phi_3 \vert^2$ can be written as the ratio of the area of subtriangle with vertices $[0, 0]$, $[\cos(h_1), \sin(h_1)]$, and $[\cos(h_2), \sin(h_2)]$ to the area of triangle with vertices $[\cos(h_3), \sin(h_3)]$, $[\cos(h_1), \sin(h_1)]$, and $[\cos(h_2), \sin(h_2)]$~\cite{Skala2008}. Area of the triangle can be calculated using the surveyor's formula~\cite{Braden1986}. Similarly, other weights can also be expressed, and the orthogonal condition given in Eq.~\ref{TOrtho1} can be expressed in terms of Cartan co-ordinates as follows.

\begin{equation*}
\sin [2(c_1 - c_3)] e^{i(d-a)\pi} + \sin [2(c_1 + c_3)] e^{i(f-b)\pi}
\end{equation*}
\begin{equation}{\label{TOrtho2}}
 - p \sin[4c_1] e^{i(g-c)\pi} = 0
\end{equation}

It can be verified that

\begin{equation}{\label{CondT1}}
- \sin [4c_1] + \sin[2(c_1 - c_3)] \geq \sin [2(c_1 + c_3)]. 
\end{equation} 

It is possible to have the following equation

\begin{equation}{\label{p}}
- p \sin [4c_1] + \sin[2(c_1 - c_3)] = \sin [2(c_1 + c_3)]. 
\end{equation}

for an appropriate value of $p$ and the orthogonal condition given in Eq.~\ref{TOrtho2} can be satisfied by choosing $(d-a)$ and $(g-c)$ as even (odd) and $(f-b)$ as odd (even). Thus, for each perfect entangler on $c_1=c_2$ plane of the Weyl chamber, there exist a pair of orthonormal product states that can be transformed into maximally entangled states. The convex hull of the squared eigenvalues of the nonlocal part of a perfect entangler existing on $c_2 = c_3$ $(c_2 = -c_3)$ plane of the Weyl chamber is also a triangle with vertices $\{e^{ih_1}=e^{ih_2}, e^{ih_3},e^{ih_4}\}$ $(\{e^{ih_1},e^{ih_2}, e^{ih_4}=e^{ih_3}\})$ and a pair of orthonormal product states that can be transformed into maximally entangled states by the nonlocal part of the perfect entangler can be constructed. 

Next, we consider perfect entanglers for which the convex hull of the squared eigenvalues of their nonlocal part is not a simplex. The argand diagram of such a perfect entangler is shown in FIG.~\ref{AD1}. The convex hull of the four points $\{1, e^{i(h_4-h_3)}, e^{i(h_1-h_3)}, e^{i(h_2-h_3)}\}$ is a quadrilateral where each point is in $k$-simplices $(k \leq 2)$ formed by the vertices of the quadrilateral~\cite{Matousek2002,Bengtsson2017}. There are four triangles formed by the vertices of the quadrilateral inscribed on the unit circle and the zero would be common to any two overlapping triangles depending on the orientations of two diagonals. In FIG.~\ref{AD1}, both $(h_1 - h_3)$ and $(h_2 - h_4)$ are less than $\pi$ or equivalently, $c_1 \pm c_3 < \pi/2$. Hence, the zero is contained in the triangle with vertices: $\{1, e^{i(h_1-h_3)}, e^{i(h_2-h_3)}\}$ and also in the triangle with vertices: $\{1, e^{i(h_4-h_3)}, e^{i(h_2-h_3)}\}$. It is possible to find two sets of co-ordinates such that 

\begin{equation}\label{CH1}
\vert \eta_3 \vert^2 e^{ih_3} + \vert \eta_1 \vert^2 e^{ih_1} + \vert \eta_2 \vert^2 e^{ih_2} = 0,
\end{equation} 
and
\begin{equation}\label{CH2}
\vert \xi_3 \vert^2 e^{ih_3} + \vert \xi_4 \vert^2 e^{ih_4} + \vert \xi_2 \vert^2 e^{ih_2} = 0,
\end{equation}
with 
$\vert \eta_3 \vert^2 + \vert \eta_1 \vert^2 + \vert \eta_2 \vert^2 = 1$ and $\vert \xi_3 \vert^2 + \vert \xi_4 \vert^2 + \vert \xi_2 \vert^2 = 1$. From Eqs.~\ref{CH1} and~\ref{CH2}, the following two product states can be constructed~\cite{Zhang2003}

\begin{eqnarray}
\vert \Phi'_1 \rangle = \vert \eta_3 \vert e^{-i(a' \pi + h_3/2)} \vert \Psi_3 \rangle + \vert \eta_1 \vert e^{-i(b' \pi + h_1/2)} \vert \Psi_1 \rangle  \nonumber
\end{eqnarray}
\begin{eqnarray}\label{PS1}
+ \vert \eta_2 \vert e^{-i(c' \pi + h_2/2)} \vert \Psi_2 \rangle,
\end{eqnarray}
and 
\begin{eqnarray}
\vert \Phi'_2 \rangle = \vert \xi_3 \vert e^{-i(d' \pi+h_3/2)} \vert \Psi_3 \rangle + \vert \xi_4 \vert e^{-i(f' \pi+h_4/2)} \vert \Psi_4 \rangle  \nonumber
\end{eqnarray}
\begin{eqnarray}\label{PS2}
+ \vert \xi_2 \vert e^{-i(g' \pi+h_2/2)} \vert \Psi_2 \rangle,
\end{eqnarray}

for some integers $a',b',c',d',f',$ and $g'$. These two product states can be converted into maximally entangled states by the nonlocal part of the perfect entangler associated with the argand diagram shown in FIG.~\ref{AD1}~\cite{Zhang2003}. The condition for these two product states to be orthogonal can be written as follows. 

\begin{equation}\label{Cond3}
\vert \eta_3 \vert \vert \xi_3 \vert e^{i(d'-a')\pi} = - \vert \eta_2 \vert \vert \xi_2 \vert e^{i(g'-c')\pi}.
\end{equation}

Squaring both sides gives the following condition.  

\begin{equation}\label{Cond3aa}
\frac{\vert \eta_3 \vert^2}{\vert \eta_2 \vert^2} = \frac{\vert \xi_2 \vert^2}{\vert \xi_3 \vert^2}.
\end{equation}

In terms of Cartan co-ordinates, this condition can be written as

\begin{equation}\label{Cond3c}
\frac{ \sin[2(c_2-c_3)] }{\sin[2(c_1+c_3)] } = \frac{\sin[2(c_2+c_3)]}{\sin[2(c_1-c_3)]}.
\end{equation}

Thus, if the above condition is satisfied by the perfect entanglers with $c_1 \pm c_3 < \pi/2$, then it is possible to construct a pair of orthonormal product states that can be converted into a pair of orthonormal maximally entangled states by those perfect entanglers. It can be observed that if $U_d(c_1, c_2, c_3)$ with $c_1 \pm c_3 < \pi/2$ satisfies the condition given in Eq.~\ref{Cond3c} then its inverse $U_d(c_1, c_2, -c_3)$ also satisfies the same condition. The condition given in Eq.~\ref{Cond3c} is valid for all perfect entanglers without any chord passing through the zero bounded by $c_1 \pm c_3 = \pi/2$, $c_1 + c_2 = \pi/2$ and $c_1 = c_2$ planes of the Weyl chamber. This region is the interior of the tetrahedron with vertices $\textnormal{CNOT}(\pi/2, 0, 0)$, $\textnormal{iSWAP}(\pi/2, \pi/2, 0)$, $\sqrt{\textnormal{SWAP}}(\pi/4, \pi/4, \pi/4)$, and $\sqrt{\textnormal{SWAP}}^{\dagger}(\pi/4, \pi/4, -\pi/4)$ [FIG.~\ref{180C}b]. 

Perfect entanglers with $c_1+c_3 > \pi/2$ and $c_1-c_3<\pi/2$ exist in the region bounded by $c_1+c_3=\pi/2$, $c_2+c_3=\pi/2$, $c_1=\pi/2$ and $c_2=c_3$ planes of the Weyl chamber. This region is a tetrahedron with vertices $\textnormal{CNOT}(\pi/2, 0, 0)$, $\textnormal{iSWAP}(\pi/2, \pi/2, 0)$, $\sqrt{\textnormal{SWAP}}(\pi/4, \pi/4, \pi/4)$, and $\textnormal{m}\sqrt{\textnormal{iSWAP}}(\pi/2, \pi/4, \pi/4)$ [FIG.~\ref{180C}b]. In this tetrahedron region, for all the perfect entanglers without any chord passing through the zero (except those on the $c_2=c_3$ plane), the zero is contained in the triangles with vertices $\{1,e^{i(h_4-h_3)}, e^{i(h_1-h_3)}\}$ and $\{1,e^{i(h_4-h_3)}, e^{i(h_2-h_3)}\}$ in their argand diagram. Two product states, which can be converted into maximally entangled states by these perfect entanglers, corresponding to these two triangles can be constructed as before and the condition for them to be orthogonal can be found as shown below. 

\begin{equation}\label{Cond3d}
\frac{ \sin[2(c_1-c_2)] }{-\sin[2(c_1+c_3)] } = \frac{-\sin[2(c_1+c_2)]}{\sin[2(c_1-c_3)] }.
\end{equation}

If $U_d(c_1, c_2, c_3)$ satisfies the conditions $c_1 + c_3 > \pi/2$ and $c_1 - c_3 < \pi/2$ then its inverse $U_d(c_1, c_2, -c_3)$ satisfies the conditions $c_1 + c_3 < \pi/2$ and $c_1 - c_3 > \pi/2$. Hence, the condition for the existence of a pair of orthonormal product state that can be converted into maximally entangled states by the perfect entanglers with $c_1+c_3 < \pi/2$ and $c_1-c_3 > \pi/2$ can be obtained from Eq.~\ref{Cond3d} by replacing $\sin[2(c_1 \pm c_3)]$ by $\sin[2(c_1 \mp c_3)]$. The condition is given below. 

\begin{equation}\label{Cond3e}
\frac{ \sin[2(c_1-c_2)] }{-\sin[2(c_1-c_3)] } = \frac{-\sin[2(c_1+c_2)]}{\sin[2(c_1+c_3)] }.
\end{equation}

These perfect entanglers are contained within the tetrahedron with vertices $\textnormal{CNOT}(\pi/2, 0, 0)$, $\textnormal{iSWAP}(\pi/2, \pi/2, 0)$, $\sqrt{\textnormal{SWAP}}^{\dagger}(\pi/4, \pi/4, -\pi/4)$, and $\textnormal{m}\sqrt{\textnormal{iSWAP}}^{\dagger}(\pi/2, \pi/4, -\pi/4)$ [FIG.~\ref{180C}b]. For all these perfect entanglers, the zero is contained in the triangles with vertices $\{1,e^{i(h_1-h_3)}, e^{i(h_2-h_3)}\}$ and $\{e^{i(h_4-h_3)}, e^{i(h_1-h_3)}, e^{i(h_2-h_3)}\}$ in their argand diagram.

\begin{figure}[h]
\begin{tabular}{c}
\includegraphics[scale=0.45]{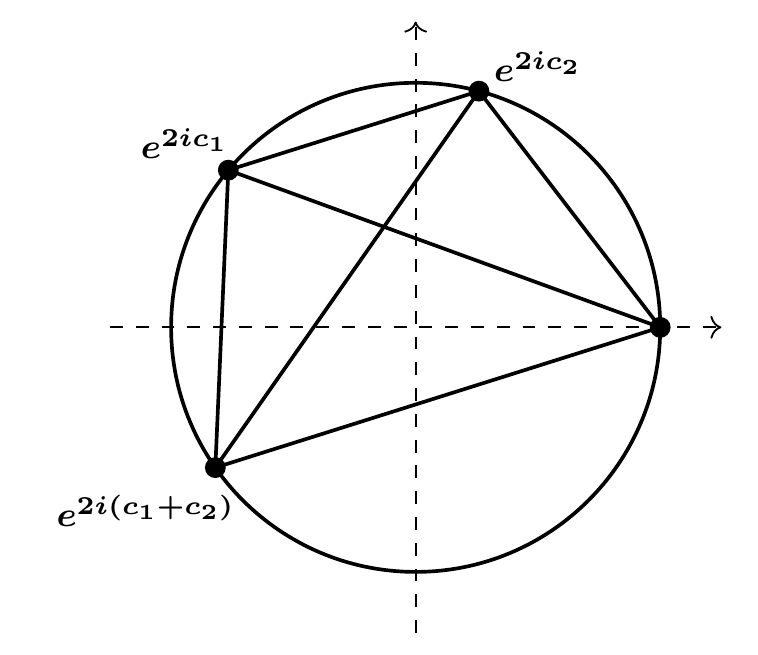} \\ (a) \\ \includegraphics[scale=0.45]{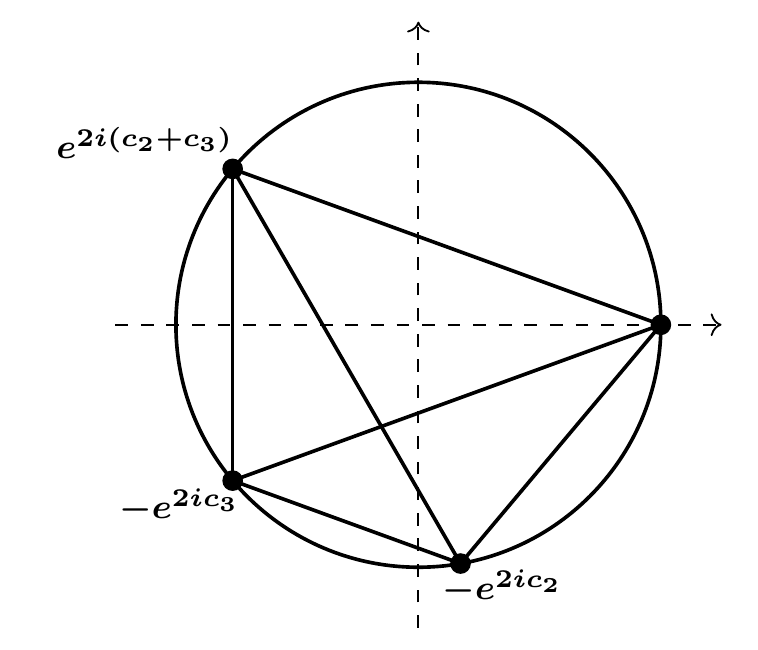} \\ (b) 
\end{tabular}
\caption{Typical argand diagrams of the nonlocal part of perfect entanglers on (a) $c_3=0$, and (b) $c_1=\pi/2$ planes of the Weyl chamber.}
\label{Trapezoid}
\end{figure}

It has to be noted that all the perfect entanglers on $c_3=0$ plane of the Weyl chamber satisfy the condition given in Eq.~\ref{Cond3c}. Similarly, all the perfect entanglers with $c_1 = \pi/2$ and $c_3 > 0$ ($c_3<0$) satisfy the condition given in Eq.~\ref{Cond3d} (Eq.~\ref{Cond3e}). Hence, all these perfect entanglers can transform a pair of orthonormal product states into maximally entangled states and their argand diagrams are shown in FIG.~\ref{Trapezoid}. 

\section{Conclusion}
To conclude, the chords in the argand diagram of the squared eigenvalues of the nonlocal part of two-qubit gates are very useful for calculating the nonlocal measures such as entangling power, gate typicality, and operator entanglement of two-qubit gates. In the argand diagram of a two-qubit gate, for the set of chords describing the entangling power, there exists another set of chords describing the gate typicality. Linear entropy of a two-qubit gate can be quantified using both sets of chords. We have studied the construction of a pair of orthonormal product states that can be transformed into maximally entangled states by the perfect entanglers using the simplices containing the zero in the argand diagram. Perfect entanglers for which the zero is contained in a $k$-simplex $(k=1,2)$ formed by the squared eigenvalues of their nonlocal part in the argand diagram can transform a pair of orthonormal product states into maximally entangled states. Perfect entanglers with at least a chord passing through the zero are represented by five different planes of the Weyl chamber. For the perfect entanglers existing on the faces of the Weyl chamber, the convex hull of the squared eigenvalues is a triangle. For the remaining perfect entanglers, the convex hull of the squared eigenvalues is a cyclic quadrilateral without any chord passing through the zero and they exist in three different tetrahedral regions of the Weyl chamber separated by $c_1 \pm c_3 = \pi/2$ planes of the Weyl chamber. For all perfect entanglers in a tetrahedral region, the zero is contained in the overlap of two specific triangles in the convex hull of the squared eigenvalues of their nonlocal part. For each tetrahedral region of perfect entanglers, we have derived a sufficient condition for the existence of a pair of orthonormal product states that can be transformed into maximally entangled states. These conditions are satisfied by the perfect entanglers existing on $c_3=0$ and $c_1=\pi/2$ planes of the Weyl chamber.  



\end{document}